# Operational resilience: concepts, design and analysis

Alexander A. Ganin[1,2], Emanuele Massaro[1,3,+], Alexander Gutfraind[4], Nicolas Steen[1], Jeffrey M. Keisler[5], Alexander Kott[6], Rami Mangoubi[7], and Igor Linkov[1,*]

[1]U.S. Army Corps of Engineers – Engineer Research and Development Center, Environmental Laboratory, Concord, MA, 01742, USA
[2]University of Virginia, Department of Systems and Information Engineering, Charlottesville, VA, 22904, USA
[3]Carnegie Mellon University, Department of Civil and Environmental Engineering, Pittsburgh, PA, 15213, USA
[4]University of Illinois at Chicago, School of Public Health, Chicago, IL, 60612, USA
[5]University of Massachusetts Boston, College of Management, Boston, MA, 02125, USA
[6]U.S. Army Research Laboratory, Network Science Division, Adelphi, MD, 20783, USA
[7]Charles Stark Draper Laboratory, Cambridge, MA, 02139, USA
[*]Igor.Linkov@usace.army.mil
[+]presently at Massachusetts Institute of Technology, Senseable City Laboratory, Cambridge, MA, 02139, USA

## ABSTRACT

Building resilience into today's complex infrastructures is critical to the daily functioning of society and its ability to withstand and recover from natural disasters, epidemics, and cyber-threats. This study proposes quantitative measures that implement the definition of engineering resilience advanced by the National Academy of Sciences. The approach is applicable across physical, information, and social domains. It evaluates the critical functionality, defined as a performance function of time set by the stakeholders. Critical functionality is a source of valuable information, such as the integrated system resilience over a time interval, and its robustness. The paper demonstrates the formulation on two classes of models: 1) multi-level directed acyclic graphs, and 2) interdependent coupled networks. For both models synthetic case studies are used to explore trends. For the first class, the approach is also applied to the Linux operating system. Results indicate that desired resilience and robustness levels are achievable by trading off different design parameters, such as redundancy, node recovery time, and backup supply available. The nonlinear relationship between network parameters and resilience levels confirms the utility of the proposed approach, which is of benefit to analysts and designers of complex systems and networks.

# Introduction

The daily functioning of modern society is necessarily challenging, and traditional risk-based approaches to managing critical infrastructure are often criticized for their inability to address widely unknown and uncertain threats[1–3]. Risk-based approaches require developing threat scenarios, evaluating system vulnerabilities and quantifying consequences associated with specific failures of system components. In the case of an unknown threat space, developing realistic scenarios proves to be an additional challenge. Moreover, it may be difficult to justify investing in hardening system components based on hypothetical and uncertain threats[4]. Weaknesses and the potentially misleading nature of risk quantification approaches, for example in cyber systems, have been pointed out by a number of researchers[5,6].

Building resilience into infrastructure networks[7] has been proposed as the key to protecting against the deleterious effect of system disruption due to natural disasters[8,9] as well as infrastructure and engineering systems' failures[10–13]. Not surprisingly, numerous interpretations of resilience have sprouted, testifying to the richness of the concept but also presenting challenges for its measurement and application[1]. Yet recent publications and guidance documents[14,15] coalesce around the definition of resilience provided by the National Academy of Sciences (NAS)[16]: Resilience of a system is its ability "to plan and prepare for, absorb, respond to, and recover from disasters and adapt to new conditions". An important feature of resilience captured in this definition is the temporal dimension: the ability to recover and retain critical system functionality in response to a wide range of threats, both known and unknown. The assessment of resilience should therefore identify the critical functionality of a system and evaluate the temporal profile of system recovery in response to adverse events. Resilience management should comparatively evaluate cross-domain alternatives designed to enhance the system's ability to (i) plan for adverse events, (ii) absorb stress, (iii) recover and (iv) predict and prepare for future stressors in order to adapt to their potential threats.

Even though definitions of resilience as a system property are commonly reported in the literature (see details in Supplement S1, table S1.1), resilience assessment has been implemented in structured but largely qualitative or semi-quantitative ways[17]. Insightful studies quantify resilience with metrics associated with different domains and subsequently integrate them into a risk-based evaluation or resilience index. For example, Bruneau et al18 identified four dimensions of seismic community resilience: technical, organizational, social, and economic. Measures of resilience – robustness, rapidity, resourcefulness, and redundancy – were then aggregated in order to minimize a function of the probability of system failures, the consequences arising from such failures, and recovery time[19]. Another qualitative but quantifiable approach sets forth a taxonomy for metrics that accommodates both change and interaction among physical, informational, and human domains[20]. The approach applies the taxonomy to cyber threats, energy systems and coastal infrastructure[9,20,21]. Such work provides insight and guidance for developing quantitative resilience measures that correspond to the qualitative identification of systemic issues and gaps. Unfortunately, it provides only limited insight into the management and control of the interconnected networks that constitute the entire system. Simultaneously, the field of network science has focused on the challenge of understanding the structure, dynamics and vulnerability of multi-layer systems across multilayer networks[22–26].

This paper proposes a methodology for quantifying a system's resilience that captures the very concept of engineering resilience advanced by the NAS[27,28] stated above. We make use of the *critical functionality* (CF) (which has been referred to before as functionality function[8], performance[29,30], quality[18,31,32]), defined as a metric of system performance set by the stakeholders, to derive an integrated measure of resilience. One example for CF, among many possible ones, is the percentage of nodes that are functioning. Another is the ratio of a network's actual flow to its maximum capacity.

We note that, in addition to resilience, CF is rich in valuable information, and can be the source of many quantitative performance metrics, such as the robustness, which we also briefly discuss. For application domains, our focus in this paper are the following two classes of models: i) multi-level directed acyclic graphs (DAG)[33], and ii) interdependent coupled networks[34]. While the second class of network is the subject of intense interest[35], the first class of networks, often overlooked by analysts, is of interest in many fields, from biology to computer science[36–38]. As an application of this class, we also approximate the Linux, specifically Ubuntu 12.04, code system, as a DAG and estimate its resilience. Because obtaining analytical results is an intractable task, even for homogeneous networks, our approach is simulation based. We do, however, obtain analytical results in the Methods section for a simple yet illuminating special case of the first model, where only nodes without redundant active supply links may be unable to supply service. This case sheds light on the relationship between redundancy and system resilience.

## Resilience: an analytical definition

A network is modeled as a graph $\boldsymbol{G(N, L)}$ with a set of nodes $\boldsymbol{N}$ connected by links $\boldsymbol{L}$. Before considering network models, the proposed concept of resilience for complex networks is described generally. The specification of $\boldsymbol{N}$ and $\boldsymbol{L}$ includes characteristics relevant to resilience, such as capacity, location, and weight of each node and link. Let $\boldsymbol{C}$ be the set of temporal decision rules and strategies to be developed in order to improve the resilience of the system during its operation. From a computational viewpoint, the parameters and algorithms defined by $\boldsymbol{C}$ depend on the particular model being implemented.

Ultimately, the system must maintain its critical functionality $K$ at each time step $t$, where $K$ maps its states or parameters to a real value between 0 and 1. This mapping may, for instance, be linear

$$K(t;\boldsymbol{N},\boldsymbol{L},\boldsymbol{C})=\frac{\sum_{i\in\{\boldsymbol{N},\boldsymbol{L}\}} w_i(t;\boldsymbol{C})\pi_i(t;\boldsymbol{C})}{\sum_{i\in\{\boldsymbol{N},\boldsymbol{L}\}} w_i(t;\boldsymbol{C})} \tag{1}$$

where $\{\boldsymbol{N}, \boldsymbol{L}\}$ is the set of all nodes and links, $w_i\,(t;\,\boldsymbol{C}) \in [0, 1]$ is a measure of the relative importance of node or link $i$ at time $t$, and $\pi_i\,(t;\,\boldsymbol{C}) \in [0, 1]$ is the degree to which a node is still active in the presence of an attack. An alternate interpretation defines $\pi_i\,(t;\,\boldsymbol{C})$ as the probability that node or link $i$ is fully functional. More complex, nonlinear and detailed definitions of critical functionality mappings are also possible. Finally, we introduce the class of adverse events (or potential attacks on targeted nodes) $\boldsymbol{E}$. For instance, in the case of a random attack on two nodes, $\boldsymbol{E}$ is the set of all attacks on all possible node pairs.

Resilience, denoted by $R$, is a composite function of the network topological properties and their temporal evolution parameters defined for a certain critical functionality and a class of adverse events $\boldsymbol{E}$:

$$R = f(\boldsymbol{N},\boldsymbol{L},\boldsymbol{C},\boldsymbol{E}) \tag{2}$$

Note that not all targeted nodes are necessarily afflicted. For a non-afflicted attacked node, we thus have $\pi_i = 1$ over the entire time interval of interest. We evaluate $R$ over a certain time interval $[0, T_C]$ where $T_C$ is the *control time*[39] which can be set *a priori*, for instance, by stakeholders or estimated as the mean time between adverse events. In continuous time, we define $R$ as

$$R \equiv R(K,\boldsymbol{E},[0,T_C]) = \frac{\frac{1}{|\boldsymbol{E}|}\sum_{\boldsymbol{E}}\int_{t=0}^{T_C} K(t;\boldsymbol{N},\boldsymbol{L},\boldsymbol{C})}{\int_{t=0}^{T_C} K^{nominal}(t;\boldsymbol{N},\boldsymbol{L},\boldsymbol{C})} \tag{3}$$

where $|\boldsymbol{E}|$ is the cardinality of set $\boldsymbol{E}$, and $K^{nominal}(t)$ is the critical functionality of the system in the case where no external events occur (Fig. 1). Equation (3) allows evaluation of the normalized dynamical performance of the system before (plan/prepare), during (absorption), and after an attack (recovery and adaptation); it intends to capture the definition advanced by the NAS given in the introduction. For computational convenience, the above equation is given in discrete time by

$$R \equiv R(K, \boldsymbol{E}, [0, T_C]) = \frac{\dfrac{1}{|\boldsymbol{E}|}\displaystyle\sum_{\boldsymbol{E}}\sum_{t=0}^{T_C} K(t; \boldsymbol{N}, \boldsymbol{L}, \boldsymbol{C})}{\displaystyle\sum_{t=0}^{T_C} K^{nominal}(t; \boldsymbol{N}, \boldsymbol{L}, \boldsymbol{C})} \tag{4}$$

In most cases, we normalize to $K^{nominal}(t) = 1$. Consequently, a normalized measure of resilience may be given by

$$R = \left(\frac{1}{T_C}\right)\frac{1}{|\boldsymbol{E}|}\sum_{\boldsymbol{E}}\sum_{t=0}^{T_C} K(t; \boldsymbol{N}, \boldsymbol{L}, \boldsymbol{C}) \tag{5}$$

We note that equation (1) embraces a large class of performance measures found in the literature. For instance, in addition to equation (5), we can also consider the measure

$$M = \min_{t \in [0, T_C]} K(t; \boldsymbol{N}, \boldsymbol{L}, \boldsymbol{C}) \tag{6}$$

The above measure is referred to as the Robustness[8,40]. Alternatively one may define $(1 - M)$ as the Risk[1].

Due to the very complex nature of networked systems and the large number of variables defining their states, it is not possible to consider all events in the set $\boldsymbol{E}$ and obtain a closed-form expression for $R$, even if all design parameters are made homogeneous across nodes, links, and time. We therefore rely on a simulation based approach. Each simulation represents a possible scenario of the networked system's evolution. For each simulation, we calculate the average value of the critical functionality $K(t,\boldsymbol{N},\boldsymbol{L},\boldsymbol{C})$ at every time step (equation 1), and from there, the resilience (equation 4 or equation 5) over the interval of interest.

The approach proposed builds upon and extends the works of others[8,18,29,30,32]. The main issue encountered when dealing with the estimation of resilience based on the simulation of the system performance curves is that those curves in the general case vary depending on the adverse events modeled. The current approach to resolve this issue is to extend the techniques of probabilistic risk analysis to resilience analysis. This extension provides the weighted average performance curve with weights representing the probabilities of the adverse events.

By contrast, in our approach, we would like to argue that the resilience of the system should not be tied to the probabilities of the adverse events to occur. Again, according to the NAS resilience is *the ability to plan, absorb, recover, and adapt*[16]. Inspired by this definition, we instead simulate the damage to the system from a certain adverse event (regardless of the probability for that event to occur) and define resilience for that particular damage. For simplicity of decision making, however, we suggest considering a certain class of adverse events. For instance, in a networked system we might define one class of adverse events as a case when functionality of 4 to 5 nodes is reduced by 40-50% (instead of defining a particular adverse event reducing the functionality of specific 4 nodes by 50%).

We illustrate the approach with two simple models: multi-level DAGs and interdependent coupled networks. We assume that *homogeneous* nodes and links comprising the network have only two possible states: *active* and *inactive*, meaning $w_i(t)$, $\pi_i(t) \in \{0, 1\}$ in equation (1). To simplify the explanation, we focus on node failures, though the concept may be extended to include links. If we denote the number of active nodes in the system at time $t$ by $A(t)$ and the total number of nodes in the system by $N$, then the critical functionality simplifies to

$$K(t) = A(t)/N \tag{7}$$

We first consider a hierarchical multi-level DAG model (Fig. 2) with $\Lambda$ levels of nodes[41,42]. We investigate how redundancy probability $p_m$, switching probability $p_s$, and the recovery time $T_R$, tradeoff parameters at the disposal of the system designer, influence the resilience of a supply-demand multi-level DAG across levels, nodes, links, and time, how they affect the absorption and recovery phases of a network's resilience profile, and how they address the optimization of network design, for a variety of attack scenarios. We also distinguish between cases where switching is instantaneous and delayed by one time step. Further description of the model is provided in the Methods (subsection 1).

Two applications of the DAG model will serve to illustrate the quantitative resilience measure introduced, as well as the method for evaluating it: 1) synthetic random hierarchical multi-level supply-demand directed acyclic graphs, and 2) the Linux, specifically Ubuntu 12.04, software network. The first is a useful, if approximate, representation of networks found in many applications[36–38,43]. The second realistically represents an existing and widely used network.

The second model is derived from the model introduced by Buldyrev et al[34] and developed by Parshani et al[44]. They consider a system comprised of two coupled undirected networks (A and B). A certain fraction of nodes in network A depends on nodes in the network B ($q_A$) and vice versa ($q_B$). If node $n$ in the network A depends on a node $m$ in the network B then node $m$ can only depend on node $n$ (or not depend on nodes in the network A at all) (see Methods, subsection 3). Without loss of generality we consider scenarios where networks A and B have the same node degree distribution. We present results for Erdos-Renyi and scale-free random networks with 800000 nodes ($N$) and average degree ($<k>$) of 2.5, and slope factor (in the scale-free case) of 2.25. Networks are generated following the algorithm presented by Catanzaro et al[45].

We consider a case with a single adverse event that destroys a number of nodes in the network. For simplicity, the adverse event happens at the time step $t = 0$. We shall refer to the result of the adverse event as the initial damage. In case of the DAG model we denote the number of nodes that become inactive (i.e., are deactivated) in level $i$ between time steps $t$ and $t + 1$ as $I_i^t$. Thus, values $I_i^0$ represent the number of nodes made inactive upon the occurrence of the adverse event. In case of the coupled networks model we denote the fraction of nodes rendered inactive in the network A as $p_{destr}$ with the assumption that the adverse event doesn't affect the network B.

## Results

### Model 1 – Directed acyclic graphs. Synthetic graphs

We consider a network composed of $N = 1000$ nodes in four levels: $N_0 = 32$, $N_1 = 87$, $N_2 = 237$, $N_3 = 644$. We first look at the special case where the switching is instantaneous with probability $p_s = 1$. Assuming the overall damage to each level is small compared to the total number of nodes in the level, the approximation derived in the Methods section (subsection 2) may be used. Figure 3 provides comparisons between the analytical calculations based on that approximation for this case and simulation results averaged over 2000 samples. We note that this case provides insight into the impact of link redundancy on both critical functionality and resilience over the time interval of interest, as we show in the first two scenarios of Fig. 4.

Examples of resilience profiles for different cases that vary in their initial damage, switching probability and recovery time are given in Fig. 4. Case 1 is a scenario where only one node in the upper level is initially disabled. This scenario represents, for instance, an accident at a power plant. It follows, that the event set $\boldsymbol{E}$ consists of all possible one-node attacks in the upper level. Critical functionality suffers minimally; it reduces from 1 to 0.97 at its lowest. Its integral, resilience, consequently suffers minimally as well: $R = 0.983$. By contrast, for a more serious attack, such as in case 2, in which five nodes at every level are disabled (such an attack might represent a large earthquake in a certain area); both critical functionality and resilience suffer. Critical functionality can be as low as 0.8 (a considerably less robust system) for a protracted number of time steps, and resilience is reduced to 0.893.

For case 3, 10 nodes are disabled, all from the top level, and the switching probability is reduced to $p_s = 0.25$. The robustness, or the critical functionality at its lowest, is more drastically reduced to 0.4, yielding an overall resilience value of 0.728. Case 4 is similar to case 3, except that the switching is delayed, i.e., if node $i$ has become disabled at time $t^*$, then the first attempt to switch is made at time $t^*$ for case 3, and at time $t^* + 1$ for case 4.

The dependency of resilience on parameters $p_m$ and $p_s$ is given in Fig. 5 (a) with the recovery time held constant at $T_R = 0.5T_C$. The figure shows that both parameters are compatible and combinable; they can be smoothly traded off to maintain a desired level of resilience. The designer here has the opportunity to select the combination of $p_m$ and $p_s$ that is least costly. Additionally, increasing $p_m$ and $p_s$ simultaneously has an observable additive effect on resilience. Beyond certain level, however, investment in redundancy yields minimal return. For instance, as shown in Fig. 5 (a), doubling the probability $p_m$ from 0.1 to 0.2 leaves the resilience unchanged for $p_s > 0.3$.

In addition, there is strong synergy between $p_m$ and $p_s$; increasing both factors together produces a rapid increase in resilience, but increasing only one or the other variable will cause the resilience metric to plateau. This can be observed in Fig. 5 (a) by regarding the resilience values shown across the phase diagram curves.

Figure 5 (b) illustrates that similar tradeoffs can be made between the maximum node recovery time $T_R$, and the switching probability $p_s$. The redundancy parameter $p_m$ is held constant at 0.01. When the recovery time is relatively short, $T_R < 0.1T_C$, resilience values close to 1 may be obtained even for values of $p_s$ as small as 0.05. Resilience is strongly affected by the recovery time, $T_R$ (Fig. 5 (b)). This temporal factor determines the characteristics of the recovery phase and has a greater impact on the calculated resilience than does the potential increase in redundancy. This is particularly true when the switching probability $p_s$ is low, as Fig. 5 (b) demonstrates.

Supplement S2 (Figures S2.1 – S2.6) displays additional results for both types of parameter dependencies. Cost and speed of design and implementation can now guide the ultimate choice from among the infinite possibilities of parameter combinations.

**Model 1 – Directed acyclic graphs. Linux software network**

The Linux software network exemplifies the structure of complex multilevel software systems and is important in its own right. This software operates in an estimated 95% of all supercomputing systems[46], and the majority of the smartphones in use (in the form of the Android operating system). Packages in Linux are linked in a formally defined hierarchy of dependencies between individual software units. In this hierarchy, a package can only be installed if all required higher level packages have previously been installed. Some redundancy is possible when multiple packages provide the same functionality. Fig. 6 shows a subnetwork of the packages network consisting of 117 nodes out of 36,902 possible nodes in the entire network. The graph data were obtained using Advanced Packaging Tool[47] on a standard installation of the Ubuntu 12.04 system.

Many modern cyber threats exploit vulnerabilities in software packages. Disabling a targeted software package leads to the failure of any services that are dependent on it. Even worse, the recovery might be protracted as a result of corrupted user data, thus requiring manual repair and cleanup. For example, an attack on the Apache web server might cause it to fail and subsequently send corrupted or maliciously designed data to backend databases[48]. Consequently, services dependent on Apache would experience data corruption, and if Apache crashes, it would be disabled as well. While the damaged server might be restarted relatively quickly, recovery from such an attack would involve checking the data, which gives rise to serious additional delays.

We evaluate the resilience of the Linux packages network in the presence of both random and guided attacks. Critical functionality and resilience profiles for guided attacks on several particularly important packages are given in Fig. 7 (a). These packages are: xauth, libstdc++6, libc6, and gcc-4.6-base. Notably in these four cases there are four sets of adverse events $\boldsymbol{E}$. Each of these sets contains only one event that successfully causes a particular node to be destructed. It is seen that the level of damage depends on which packages are targeted.

For random attacks (Fig. 7 (b)), we consider another set of adverse events, consisting of 36,902 events. In this case, resilience is significantly higher than in the case of guided attacks due to the low importance of many of the nodes that failed from the attacks, thus yielding $R = 0.99975$ and $M = 0.999$.

**Model 2 – Interdependent networks. Synthetic graphs**

We summarize the results for the second model in the Fig. 8. Panels (a) and (b) show the dependency of the critical functionality of a system of two interdependent Erdos-Renyi (ER) networks on time for 2 distinct cases of the recovery resources available expressed as the number of the backup agents ($N_b$). As it is evidenced by the Fig. 8 depending on the value $N_b$ there is a sharp distinction between two cases: in Fig. 8 (a) the system is unable to return to the recovered state and the critical functionality oscillates between 0 and about 0.5; due to random duration of cascading recovery and failure processes eventually the amplitude of the oscillations of the mean value of CF decreases; in Fig. 8 (b) the backup supply to $0.4N$ nodes in the network A allows reaching a stable state after the backup supply removed and further recovery of the system.

In scale-free networks (SF) the results for $\langle k \rangle = 2.5$ and $N = 800000$ (Fig. 8 (d), (e)) show a much larger dispersion than in Erdos-Renyi networks. In particular some of those networks suffer a much smaller drop in critical functionality in response to the adverse events modeled. This obviously is a consequence of the infinite dispersion in degrees distribution in the SF networks (though in our case the dispersion is finite due to the finite number of nodes). Another distinct specific trait of the SF networks with small value of the average degree is the fact that whether the network fully recovers or not is strongly dependent on the stochastic nature of the cascading failure process. Particularly it is obvious from Fig. 8 (d) that the success of recovery is determined by whether the most important hubs were affected during the deactivation process. If those hubs are not affected the damage is relatively small, otherwise the damage results in large drop in critical functionality and no full recovery within the control time.

Finally panels (c) and (f) in the Fig. 8 show the phase diagrams of the dependency of the resilience (calculated as the integral of the CF over the control time) on the model parameters $p_{destr}$ and $N_b$. Notably the critical functionality practically drops to zero for only 0.2 of the nodes destroyed in the network A initially ($p_{destr}$). Parshani et al[44] has demonstrated that if $p_{destr}$ is more than 0.2545 the network experiences the first order transition leading to a state with almost no active nodes. We have reproduced their results for the Erdos-Renyi networks and confirmed that if $p_{destr}$ is less than the threshold of 0.2545 the transition doesn't occur. However due to minor modifications we made to the network generation algorithms aimed at connecting all the nodes in a single giant component (GC) in the beginning of the process, we observe decrease of the threshold value $p_{destr}$, causing the first order transition, to about 0.15 – 0.2. After the drop of the critical functionality (due to the cascading failure) on the step $T_R$ the recovery process starts. The recovery is successful only if $N_b$ is about $0.4N$ or higher. Finally if the whole network A is destroyed as a result of the adverse event ($p_{destr} = 1$) then the recovery cannot start due to the absence of the GC (A). Results for the scale-free network show similar tendencies although notably are much more disperse in the region $N_b \in [0.1; 0.6], p_{destr} \in [0.05; 0.85]$. We interpret this as the consequence of the divergence of the standard deviation of the degree distribution in the scale-free networks with $\gamma \in (2; 3)$.

## Conclusion

We have presented a detailed approach for implementing the NAS definition of resilience as a function of design tradeoff parameters, as illustrated in the study with multi-level directed acyclic graphs and interdependent networks. The approach allows evaluation of resilience across time, and not just as a single quantity. Designers can thus analyze the effect of parameter choice and design emendations on overall network resilience and robustness. Focusing on multi-level directed acyclic graphs and interdependent networks, we have demonstrated how network parameters can be traded off to obtain a desired resilience and other performance measures' level. Future work will extend to multiplex systems, other real life networks. An important long term challenge is to model adaptation, which is part of the response cycle that follows restoration and includes all activities that enable the system to better resist similar adverse events in the future.

## Methods

### Absorption and recovery algorithms in the DAG model

A hierarchical multi-level DAG (Fig. 2) has $\Lambda$ levels of nodes[41,42]. Each layer is comprised of $N_i$ nodes ($i = 0, \ldots, \Lambda - 1$). The links represent a supply–demand relationship. A link starts at a supplier node and ends at a demander node. Thus, every link in the network is directed. For every level, we identify a set of services that all nodes in a particular level supply. Then, for every service supplied in the network, we define whether or not it depends on other services and a (possibly empty) list of the dependencies. The levels are ordered in such a way that links only go from a *higher* level $i$ to a *lower* level $j$. With this convention, $i < j$, or, the higher the level, the smaller the index. Additionally, no links can be formed between nodes in the same level. We also disallow cycles in the network by imposing the following constraint: a node cannot depend on any of the services provided by any of the other nodes in its level or on any of the services provided in any of the lower levels. Initially, all dependencies are resolved, and every node has one incoming link from one or more upper levels on which it depends for the supply of its services. Furthermore, for every dependent node and each of its required services, we introduce a list of potential suppliers. The probability that a node has a link from each of the potential suppliers is $p_m$. Said another way, a node has many links supplying a given service but only one of those links is enabled initially (known as *real*), while the others are contingent backup links (known as *virtual*), should they exist.

To model an adverse event, we introduce an ability to destruct a node for a time period $T_R$, as was recently done by Majdandzic et al[39]. A *destructed* node is inactive and is therefore unable to supply services until it recovers. Another possible cause of deactivation is an unresolved dependency, that is, the absence of a real link to a node supplying a required service. This can happen when the only supply node available for a service is either destructed or its upstream supplier is destructed. We shall refer to the nodes with an unresolved dependency as *disabled* nodes. Note that a node can be disabled, destructed, or both at a given time. Once a node becomes inactive, all of its dependencies, connected through their real links, are subject to deactivation unless they have other real links providing all of their required services (Fig. 2).

We assume that a node is eligible to *switch* links, that is, to turn a virtual or contingent link into a real one, if virtual links for all of the node's unresolved dependencies exist and the node is not destructed. At every time step during which the node is both disabled and eligible to switch, switching happens with probability $p_s$. Switching can be either *instant* (the first attempt to switch is made at the same time step the node has become disabled) or *delayed* (the node with an unresolved dependency remains disabled for at least one time step).

## Analytical approximation for the special case of the DAG model

In this section we derive equations describing the number of active nodes in the special case where the switching is instantaneous with probability $p_s = 1$ while the initial damage is small compared to the total number of nodes in the network.

Let us denote by $\Lambda$ the number of levels in the network and by $N_i$ the number of nodes in level $i$ ($i = 0, \ldots, \Lambda - 1$). We can find the probability that a node in level $i$ has only one service provider in level $j$ as follows:

$$f_j = (1 - p_m)^{(N_j - 1)} \tag{8}$$

For the case where the number of deactivated nodes at each time step is small enough or in which $p_m = 0$, we may assume that only the nodes with one link for a relevant service are disabled as a result of the inactivation of their supplier (thus neglecting the cases in which the node has more than one supplier of a service and all of them are deactivated).

The average number of active nodes in level $i$ at time step $t$ ($A_i^t$) is given by the formula:

$$A_i^t = N_i - \sum_{s=0}^{t-1} I_i^s \tag{9}$$

These $A_i^t$ nodes have $A_j^t$ suppliers in each level $j < i$. We disabled $I_j^{t-1}$ suppliers of the nodes in level $j$ between steps $t - 1$ and $t$. The probability for a node in level $i$ to become disabled between time steps $t$ and $t + 1$ is therefore:

$$1 - \prod_{j=0}^{i-1} \left( 1 - {I_j^{t-1} f_j} \big/ {A_j^{t-1}} \right) \tag{10}$$

We approximate the distribution of $1 / A_j^{t-1}$ to be linear in $I_i^s$, although the dependency itself is not linear:

$$\frac{1}{A_j^{t-1}} = \frac{1}{N_j - \sum_{s=0}^{t-1} I_j^s} = \frac{1}{N_j}\left(\frac{1}{1-x}\right) \tag{11}$$

Where (under the assumption that the overall damage is small enough):

$$x = \frac{\sum_{s=0}^{t-1} I_j^s}{N_j} << 1 \tag{12}$$

Considering the Taylor expansion of $1 / (1-x)$, we have for small values of $x$: $1 / (1-x) \approx 1 + x$.

Then, on average between steps $t$ and $t + 1$, we disable the following number of nodes in level $i$:

$$I_i^t = \left\{1 - \prod_{j=0}^{i-1}\left[1 - \left(I_j^{t-1} f_j \Big/ \left(N_j - \sum_{s=0}^{t-2} I_j^s\right)\right)\right]\right\}\left(N_i - \sum_{s=0}^{t-1} I_i^s\right) \tag{13}$$

After the recovery period of $T_R$ time steps has transpired, the initially destroyed nodes are rebuilt and become active unless they still lack sufficient supplies from the upper levels. Thus, assuming that $T_R > \Lambda$, $I_i^t = 0$ for $i = \{0, \ldots, \Lambda\}$ and $t = \{\Lambda, \ldots, T_R - 1\}$.

The total number of nodes restored at step $T_R$ in level $i$ is given by the expression:

$$\left|I_i^{T_R}\right| = \frac{\left|I_i^0\right|}{N_i}\left(N_i - F_i^{T_R}\right) \tag{14}$$

Here, $F_i^t$ represents the total number of nodes disabled in level $i$ at time step $t$ due to the fact that they do not have sufficient supplies from the upper levels:

$$F_i^t = \left\{1 - \prod_{j=0}^{i-1}\left[1 - \frac{\left(N_j - A_j^{t-1}\right)}{N_j}\right]\right\} N_i \tag{15}$$

And for the next steps, the formula is as follows:

$$\left|I_i^t\right| = \left\{1 - \prod_{j=0}^{i-1}\left(1 - \frac{\left|I_j^{t-1}\right| f_j}{A_j^{t-1}}\right)\right\}\left(N_i - F_i^t\right) \quad (16)$$

Using the formulae above, we may evaluate the average approximated resilience profiles and find the values of resilience.

**Absorption and recovery algorithms in the coupled networks model**

The failure propagation algorithms are described in the original model of Buldyrev et al[34]. Initial damage results in a certain fraction of nodes deactivated in the network A. Once those nodes are deactivated the network A is fractured in clusters. Nodes that do not belong to the largest cluster of the network A are also assumed deactivated. Then all the nodes in the network B that depend on the deactivated nodes in the network A are also deactivated. It results in fracturing of the network B, and the nodes that are not in the largest cluster of the network B are also assumed deactivated. In the second step of the process nodes in the network A depending on the deactivated nodes in the network B are deactivated and the process propagates in the same fashion until there are no more nodes to deactivate in any of the networks.

Recovery is accomplished by the backup supply agents replacing unresolved dependencies of the nodes in the first network (A). The number of those agents is denoted $N_b$. Each backup agent can serve only one node at a time. Nodes to provide the backup supply to are chosen randomly from those nodes in the network A that depend on a currently inactive node in the network B. Thus backup is provided either to all nodes in the network A with an unresolved dependency (in this case full recovery is guaranteed) or to $N_b$ nodes only. If a node has backup supply and it is connected to its network's GC it is activated. Once a node is activated it is included in the network's GC. This causes eventual growth of the giant component of the network A. After that the nodes in the network B that depend on the activated nodes in the network A and are connected to the GC of the network B are also activated and the process propagates in a similar fashion. Once the process is complete the recovery phase finishes. After that the backup supply is removed meaning that all the nodes whose supplier in the network B is not active (after the recovery phase) are deactivated. This leads to a cascading failure propagating as described in the introduction section. Once the failure phase finishes the recovery phase is repeated until the full network recovery is established or the control time is reached.

Let us consider a simple two-network system (Fig. 9). At the beginning of simulation (time 0) two nodes (A1 and A3) are assigned to be initially destroyed. Cascading process finishes before the recovery time (that is time to repair a node is more than the cascading failure time) $T_R$. Thus at time $T_R$ the network is in the state it had after the cascading failure. After $T_R$ steps have passed the recovery process starts. The case of $N_b = 0$ is shown in the panel (a) of the Fig. 9. In this case the only recoverable node is A3. After its recovery node B3 is also recovered, but further recovery is not possible. Nodes A1, A2, B1, and B2 can't be recovered as they have an unresolved dependency. In addition even if nodes A1, B1 were independent they still wouldn't have been recoverable due to the fact that they are not connected to the respective networks' GCs.

Now consider the case $N_b = 1$. In this case two scenarios are possible:

a) The node chosen for backup supply is A1 (Fig. 9 (b)). Then no recovery can happen as this node is not connected to the network A GC (or GC(A)) and the recovery phase ends in 0 steps;

b) The node chosen for backup supply is A2 (Fig. 9 (c)). Then this node recovers, node B2 recovers in turn. During the second step of the recovery phase node A1 recovers and node B1 recovers in turn.

## Author contributions statement

All authors developed the concept and the model, A.G. developed the software and conducted the experiment, A.G., E.M., A.Gu., I.L. analyzed the results, I.L. conceived the idea and provided overall guidance. N.S., J.K., A.K., R.M. reviewed the manuscript. This study was funded by the US Army. Permission was granted by the USACE Chief of Engineers to publish this material. The views and opinions expressed in this paper are those of the individual authors and not those of the US Army or other sponsor organizations.

## Additional information

The author(s) declare no competing financial interests.

## Figure legends

**Figure 1.** *Resilience and critical functionality concepts as advanced by the NAS.* The system's resilience is evaluated as the integral of the critical functionality's ($K$) dependency on time.

**Figure 2.** *Network generation and modeling of an adverse event.* The hierarchical area is first defined, then links are established according to the Bernoulli trial probability law, with parameter $p_m$. During the operation, if a node with a redundant link or links is made inactive, it can switch with probability $p_s$ at each time step following attack time $t$. After the repair time period of $T_R$ steps elapses following the attack, the initially destroyed nodes are restored.

**Figure 3.** *Special case: analytical vs computational solution.* Comparison between the analytical solution and the computational experiments in the limiting but insightful special case in which switching is instantaneous when an additional link is available, meaning that $p_s = 1$, under three different scenarios. Initial damage numbers for each layer are ordered as follows: $I_0^0, I_1^0, I_2^0, I_3^0$. For instance, the initial damage in scenario 1 is: $I_0^0 = 1, I_1^0 = I_2^0 = I_3^0 = 0$. This special case reveals the impact on resilience of redundancy levels, as represented by the probability $p_m$. The cylinder represents the 25–75 percentile range.

**Figure 4.** *Resilience profiles for different scenarios in synthetic graphs.* Results are shown for the redundancy probability parameter $p_m$ = 0.01. Initial damage numbers for each layer are ordered as follows: $I_0^0, I_1^0, I_2^0, I_3^0$. For instance, the initial damage in scenario 1 is: $I_0^0 = 1, I_1^0 = I_2^0 = I_3^0 = 0$. The robustness (*M*) values for scenarios 1 – 4 is the minimum value for each curve: 0.966, 0.787, 0.453, and 0.395 respectively.

**Figure 5.** *Resilience as a function of design parameters.* (**a**) Resilience (value shown on curves, σ ∈ [3.1E-3; 2.0E-2]) dependencies on switching probability at each time step, or $p_s$, and redundancy parameter $p_m$, for a four level hierarchical network where the initial number of destroyed nodes at each level is $I_0^0 = 16$, $I_1^0 = I_2^0 = I_3^0 = 0$ respectively, and recovery time is held constant at $T_R = 0.5T_C$, and (**b**) resilience (σ ∈ [5.9E-4; 4.3E-2]) dependencies on $p_s$ and $T_R$ for $I_0^0 = I_1^0 = I_2^0 = I_3^0 = 5$ with constant $p_m$ = 0.01 (color bar indicates the value of the resilience).

**Figure 6.** *Subnetwork of the Linux hierarchical packages network.*

**Figure 7.** *Resilience profiles for the Linux network.* (**a**) Guided attacks, and (**b**) random attacks. It is clear that guided attacks are considerably more damaging. Moreover, not all guided attacks are equally damaging; as shown in (**a**), attacks on xauth are less damaging than on libstdc++6. Most damaging are attacks on libc6 and gcc-4.6-base. The robustness (*M*) values for scenarios 1 – 4 in the panel (**a**) are 0.982, 0.655, 0.130, and 0.129 respectively, while for the case in the panel (**b**) *M* = 0.999.

**Figure 8**. *Representative profiles of the dynamics of K and resilience in networks with N = 800000.* Panels (**a**) and (**b**) show results for ER networks with $p_{destr} = 0.5$, $N_b = 0.35N$ and $N_b = 0.4N$ respectively. Panels (**d**) and (**e**) show results for SF networks with $p_{destr} = 0.5$, $N_b = 0.1N$ and $N_b = 0.62N$ respectively. In the panels (**a**), (**b**), (**d**), (**e**) the solid line corresponds to the mean value of $K$ over 100 simulations. Gray area corresponds to the region $K \pm \sigma(K)$ (where σ is the standard deviation). The plot in the panel (**d**) also shows simulations where critical functionality restores to 1. It follows that the success of the restoration algorithm depends for the most part only on the results of the cascading failure (rather than the random selection of the nodes with the backup supply): in cases when the important hubs are active after the cascading failure recovery is more likely. Finally panels (**c**) and (**f**) display phase diagrams of the resilience dependencies on both $N_b$ and $p_{destr}$ parameters. In both ER and SF networks the recovery process is stochastic and very sensitive to the backup supply available.

**Figure 9.** *Recovery process in coupled networks.* At time 0 nodes A1 and A3 are deactivated for $T_R$ steps. The failure process finishes before $T_R$ so at time $T_R$ the network is in its state after the cascade. Panel (**a**) illustrates the case $N_b = 0$. Only independent node A3 can be recovered as it is connected to the network A GC. Once node A3 is activated its dependent node B3 is also activated. Panels (**b**) and (**c**) illustrate the stochastic nature of the recovery process. In these cases $N_b = 1$. At time $T_R + 1$ either node A1 (**b**) or node A2 (**c**) can have backup supply. In case (**b**) the recovery phase doesn't start. After that backup is removed but cascading failure occurs. On the next time step when backup is applied again to a randomly selected node the recovery cascade is possible if the node chosen is A2. This case is the same as the case (**c**) with time $T_R + 1$ (in the case (**c**)) corresponding to the time $T_R + 3$ (in the case (**b**)).

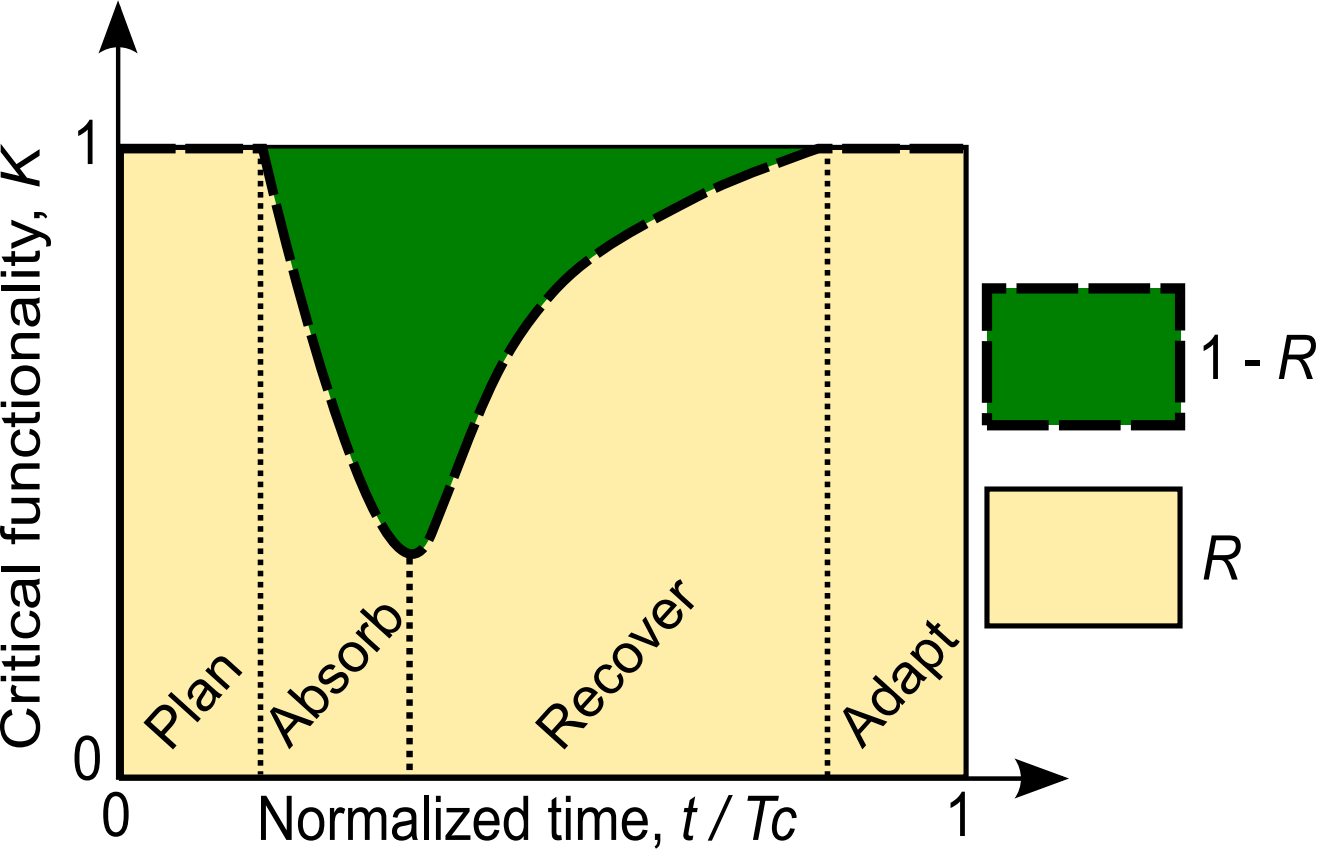

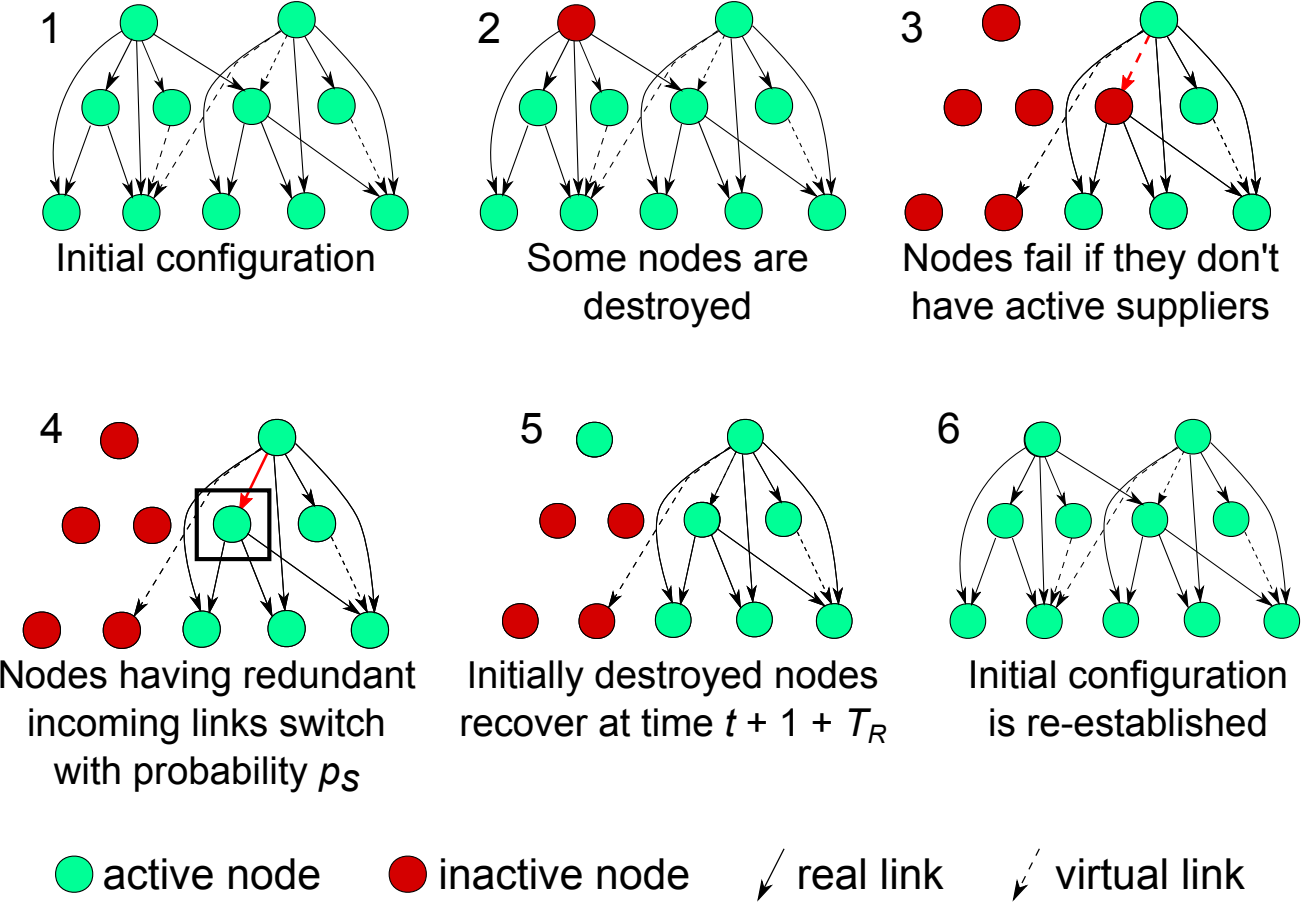
1
Initial configuration
2
Some nodes are destroyed
3
Nodes fail if they don't have active suppliers
4
Nodes having redundant incoming links switch with probability $p_S$
5
Initially destroyed nodes recover at time $t + 1 + T_R$
6
Initial configuration is re-established
active node
inactive node
real link
virtual link

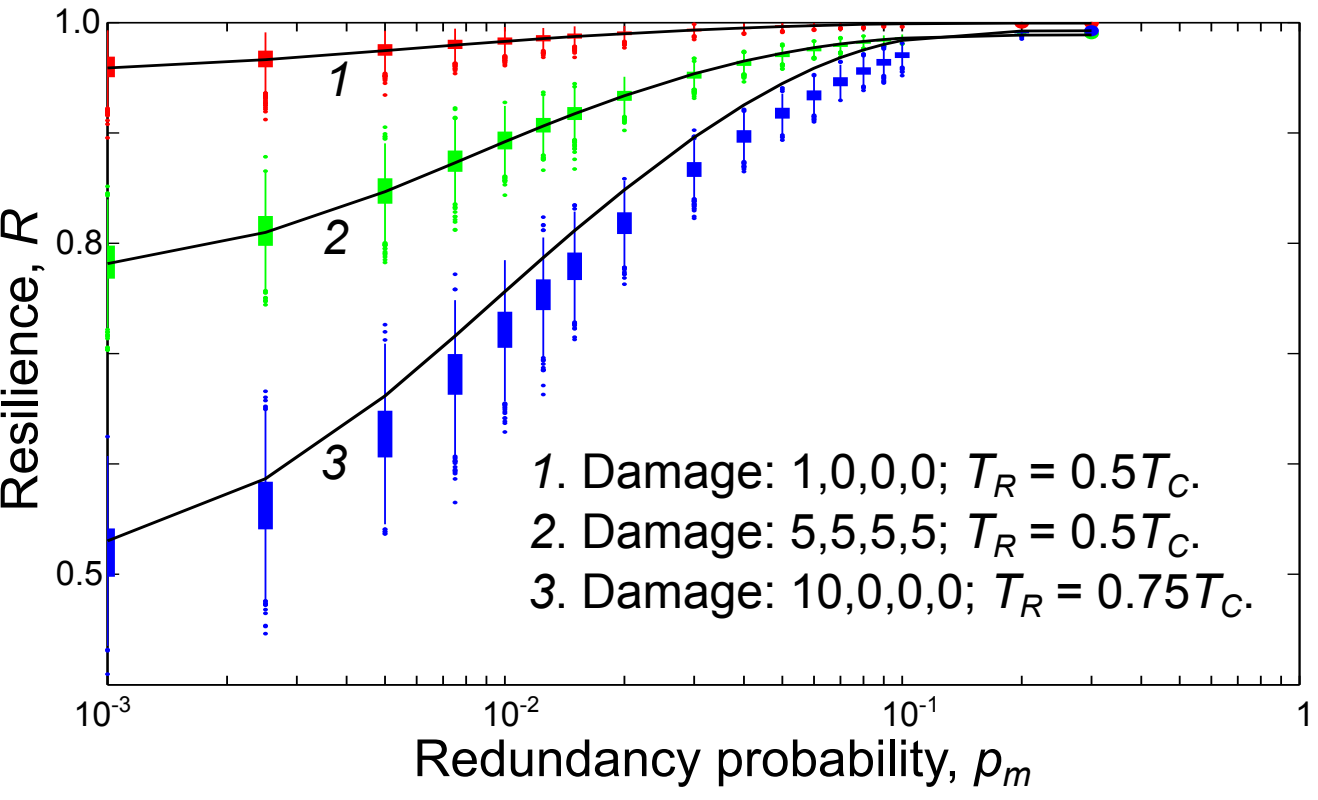

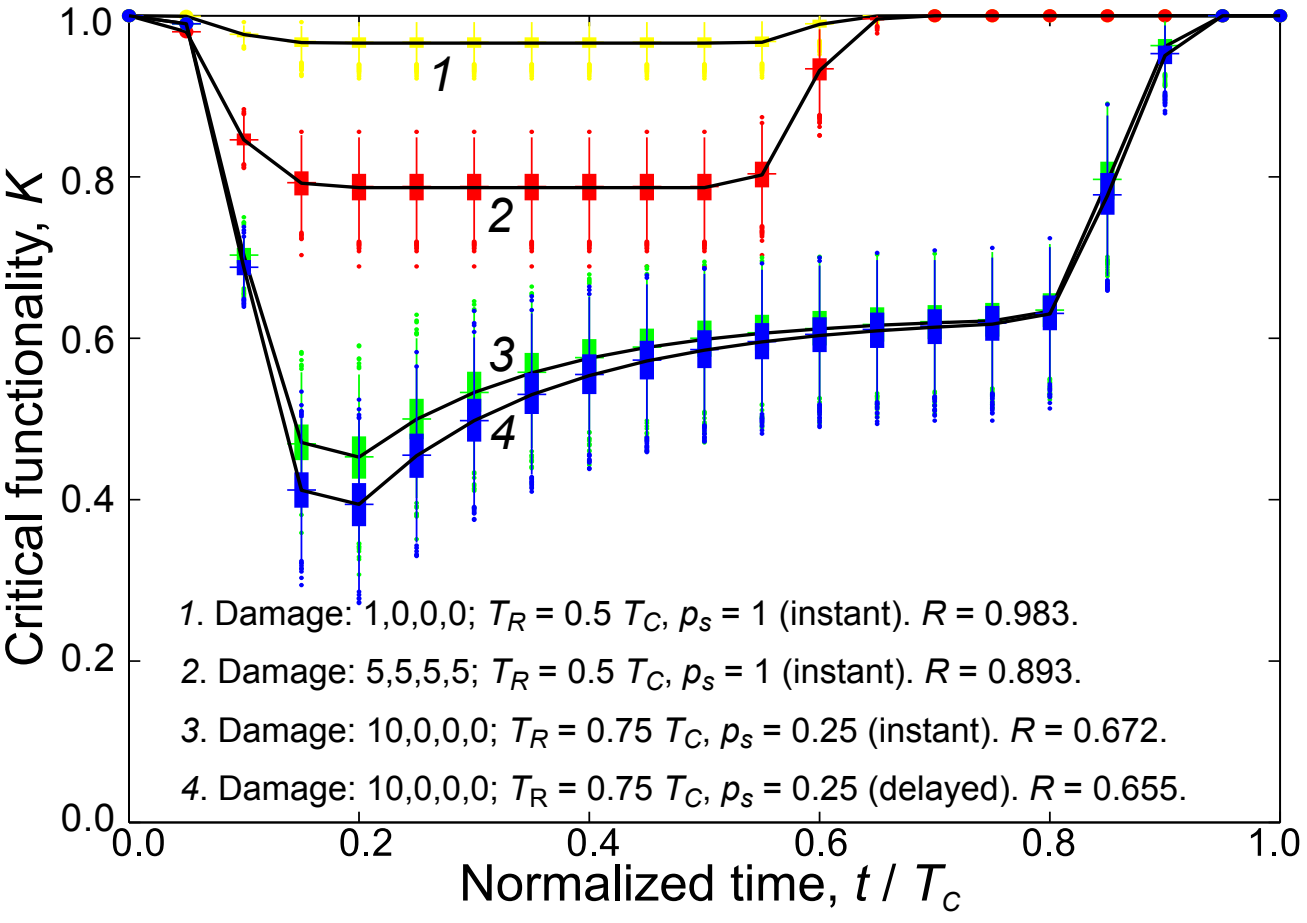

1.0
0.8
0.6
0.4
0.2
0.0
Critical functionality, K
0.0
0.2
0.4
0.6
0.8
1.0
Normalized time, t / T_C
1
2
3
4
1. Damage: 1,0,0,0; T_R = 0.5 T_C, p_s = 1 (instant). R = 0.983.
2. Damage: 5,5,5,5; T_R = 0.5 T_C, p_s = 1 (instant). R = 0.893.
3. Damage: 10,0,0,0; T_R = 0.75 T_C, p_s = 0.25 (instant). R = 0.672.
4. Damage: 10,0,0,0; T_R = 0.75 T_C, p_s = 0.25 (delayed). R = 0.655.

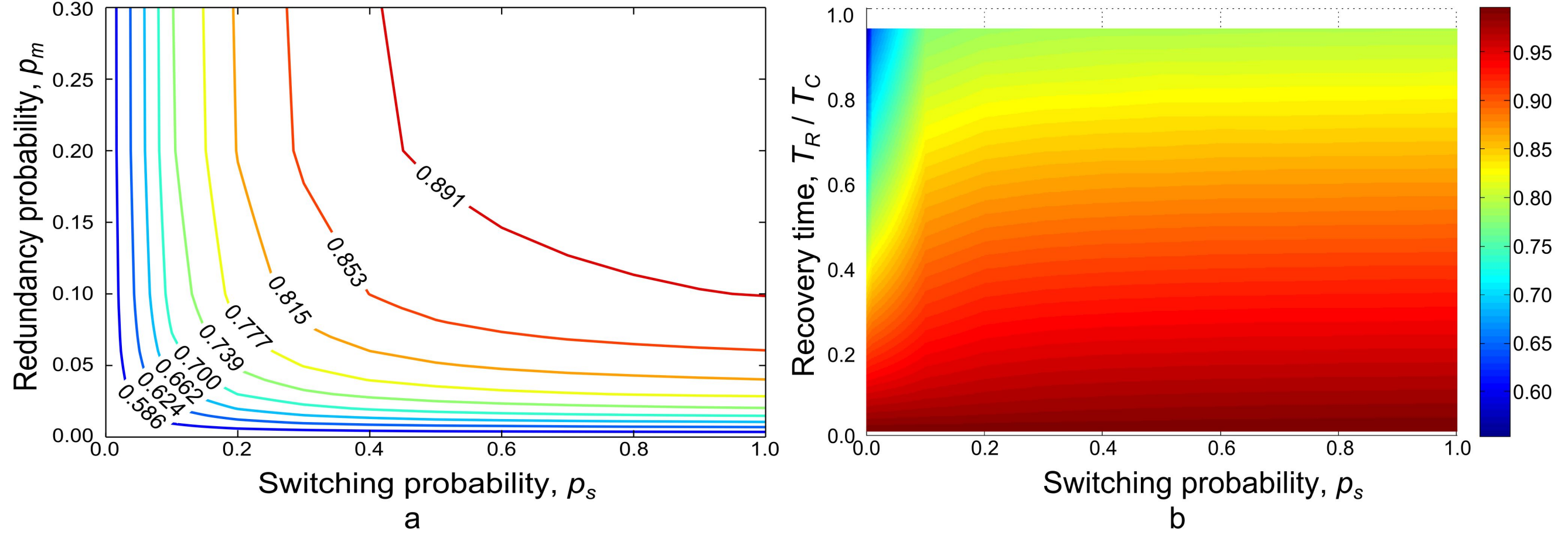

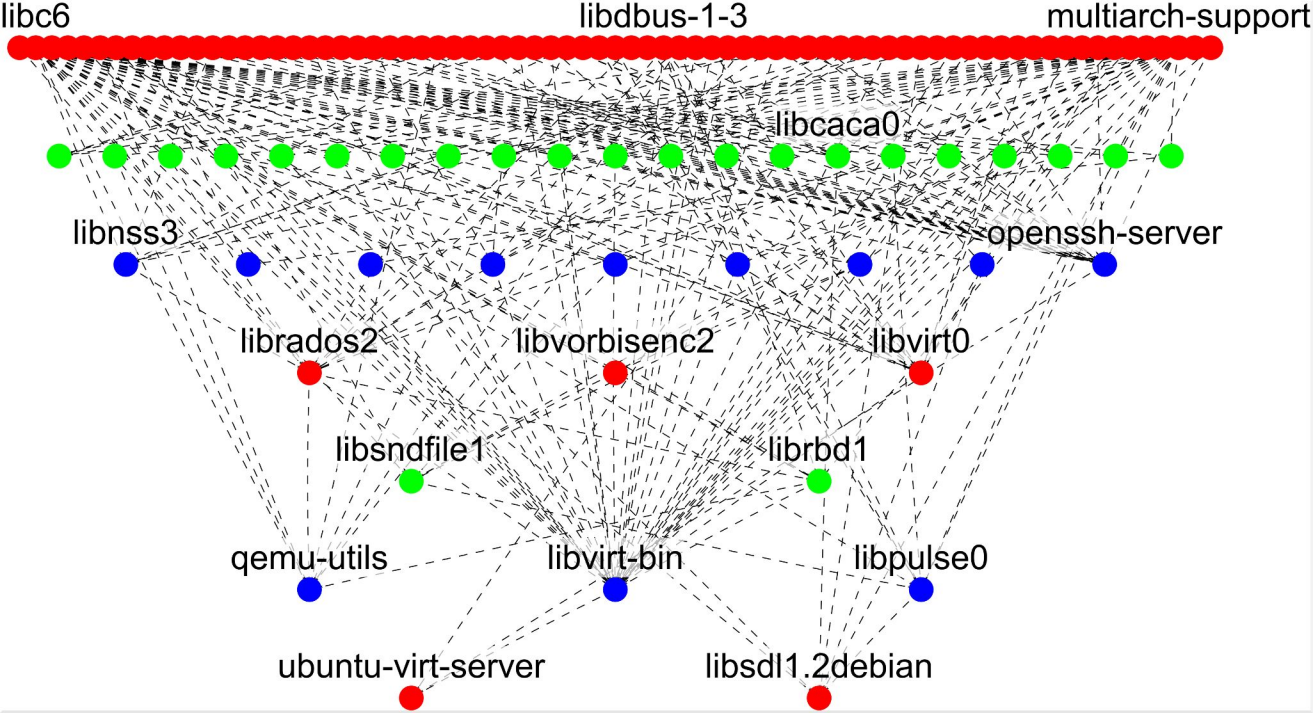

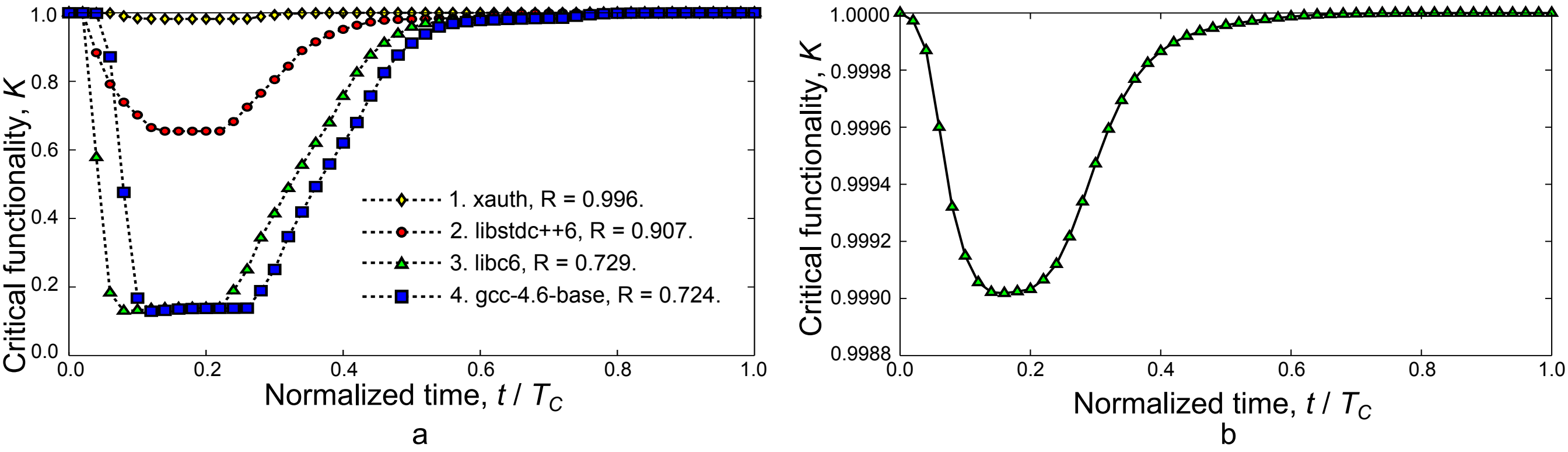

Critical functionality, K
Normalized time, t / T_C
1. xauth, R = 0.996.
2. libstdc++6, R = 0.907.
3. libc6, R = 0.729.
4. gcc-4.6-base, R = 0.724.
a
Critical functionality, K
Normalized time, t / T_C
b

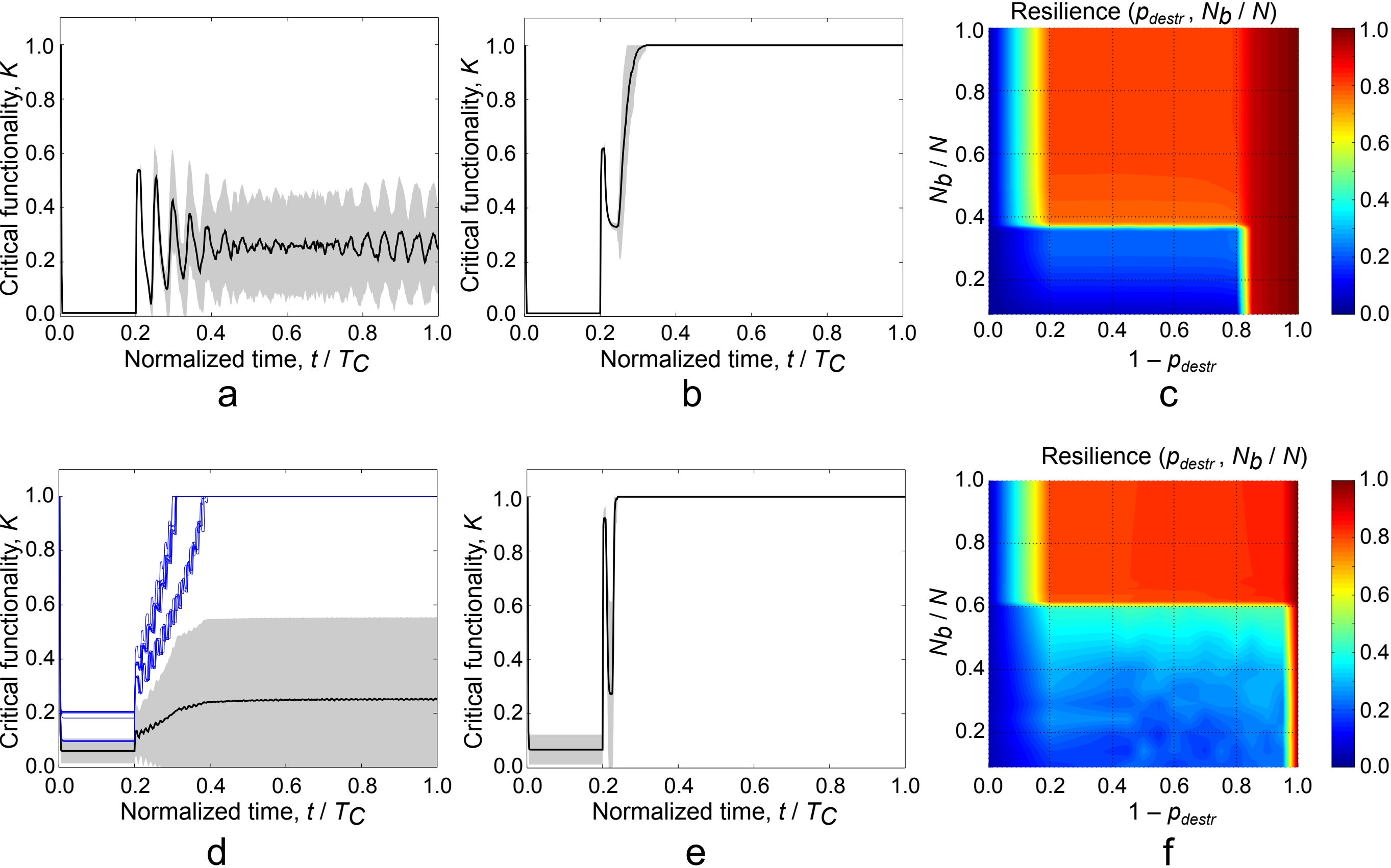

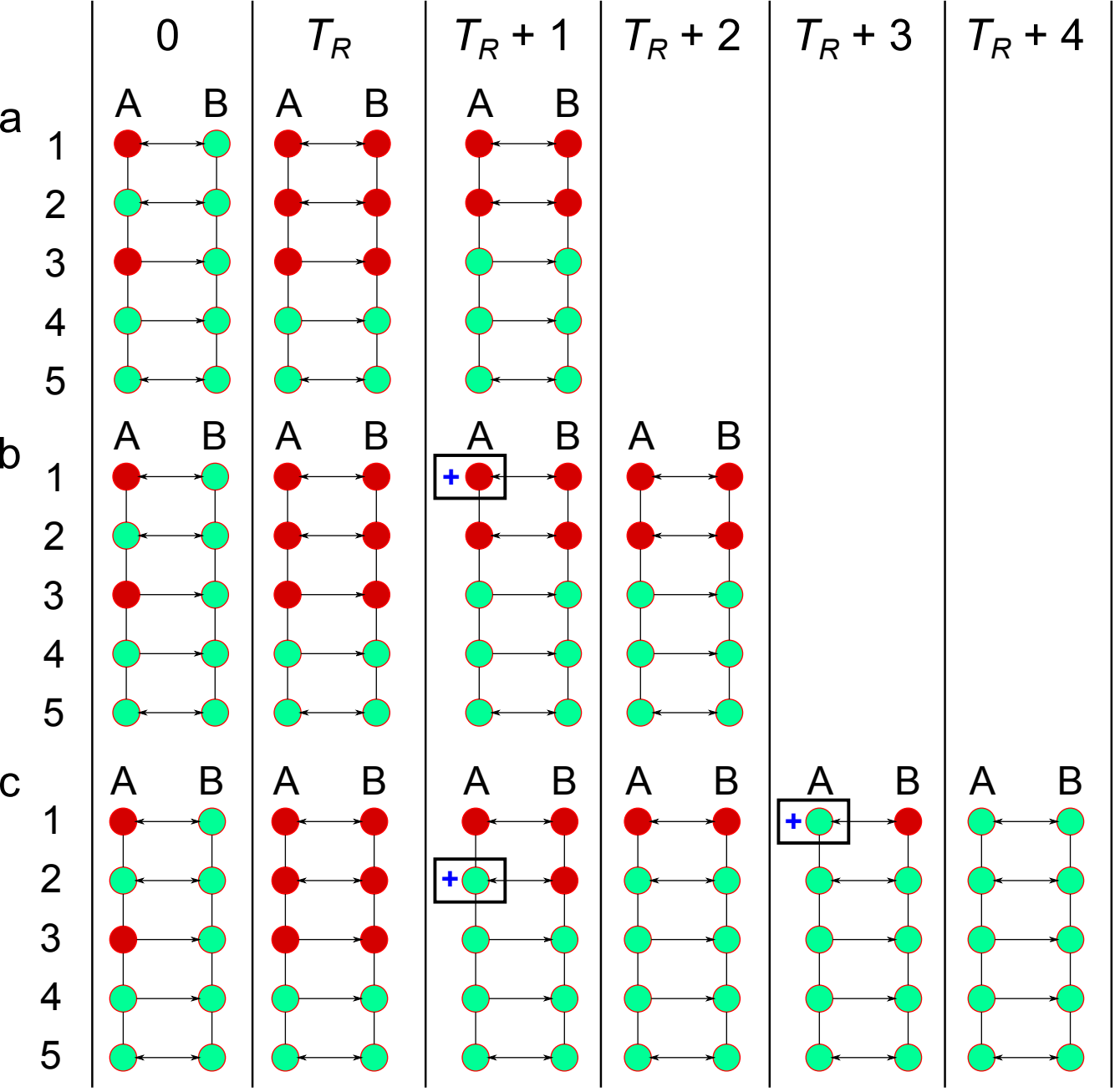